\documentclass[journal]{IEEEtran}
\usepackage{amsmath,amsfonts}
\usepackage{array}
\usepackage{textcomp}
\usepackage{stfloats}
\usepackage{url}
\usepackage{verbatim}
\usepackage{graphicx}
\def\BibTeX{{\rm B\kern-.05em{\sc i\kern-.025em b}\kern-.08em
		T\kern-.1667em\lower.7ex\hbox{E}\kern-.125emX}}
\usepackage{balance}
\usepackage[numbers, sort]{natbib}
\usepackage{amssymb,mathtools,bm}
\usepackage{algorithm,algpseudocode}
\usepackage{booktabs,multirow}
\usepackage{tabularx}
\newcolumntype{Y}{>{\centering\arraybackslash}X}
\usepackage{threeparttable}
\usepackage{xcolor}
\usepackage{enumerate}
\graphicspath{{pic/}}
\usepackage{subcaption}
\usepackage{bbding}
\usepackage{acronym}
\usepackage{etoolbox}
\usepackage{nomencl}
\definecolor{revise_blue}{rgb}{0.00,0.00,0.00}
\makenomenclature
\renewcommand{\nomgroup}[1]{%
	\item[\textit{%
		\ifthenelse{\equal{#1}{A}}{Abbreviations}{}%
		\ifthenelse{\equal{#1}{B}}{Indices and Sets}{}%
		\ifthenelse{\equal{#1}{C}}{Variables}{}%
		\ifthenelse{\equal{#1}{D}}{Parameters and Constants}{}%
	}]%
}

\makeatletter
\patchcmd{\NAT@test}{\else\NAT@nm}{\else\NAT@nmfmt{\NAT@nm}}{}{}
\let\NAT@up\itshape
\makeatother
\newcommand{\ith}{$i^\mathrm{th}$ }
\newcommand{\onecolfigure}[2]{
	\begin{figure}[!htb]
		\centering
		\includegraphics[scale=1]{#1}
		\caption{#2}
		\label{fig:#1}
	\end{figure}
}

\begin{document}

\title{Adaptive Multi-Objective Bayesian Optimization for Capacity Planning of Hybrid Heat Sources in Electric-Heat Coupling Systems of Cold Regions}

\author{Ruizhe Yang,  Zhongkai Yi,~\IEEEmembership{Member,~IEEE,} Ying Xu,~\IEEEmembership{ Senior Member,~IEEE,} Guiyu Chen, Haojie Yang, Rong Yi, Tongqing Li, Miaozhe Shen,
    Jin Li, Haoxiang Gao, Hongyu Duan,
    \thanks{Manuscript received 31 September 2024; revised 18 December 2024; accepted 6 January 2025. This work is supported by the National Natural Science Foundation of China under Grant No. 52407089. (\textit{Corresponding Author: Zhongkai Yi and Ying Xu.} e-mail: yzk@hit.edu.cn; ying.xu@hit.edu.cn.)}
    \thanks{Ruizhe Yang, Zhongkai Yi, Ying Xu, Guiyu Chen, and Haojie Yang are with the School of Electrical Engineering and Automation, Harbin Institute of Technology, Harbin 150006, China.} %
    \thanks{Rong Yi, Tongqing Li, Miaozhe Shen, Jin Li, Haoxiang Gao, and Hongyu Duan are with the State Grid Heilongjiang Electric Power Co., Harbin 150006, China.} %
}

\markboth{IEEE TRANSACTIONS ON INDUSTRY APPLICATIONS, ~Vol.~X, No.~X, January ~2025}%
{Shell \MakeLowercase{\textit{et al.}}: A Sample Article Using IEEEtran.cls for IEEE Journals}


\maketitle

\begin{abstract}
    The traditional heat-load generation pattern of combined heat and power generators has become a problem leading to renewable energy source (RES) power curtailment in cold regions, motivating the proposal of a planning model for alternative heat sources. The model aims to identify non-dominant capacity allocation schemes for heat pumps, thermal energy storage, electric boilers, and combined storage heaters to construct a Pareto front, considering both economic and sustainable objectives. The integration of various heat sources from both generation and consumption sides enhances flexibility in utilization. The study introduces a novel optimization algorithm, the adaptive multi-objective Bayesian optimization (AMBO). Compared to other widely used multi-objective optimization algorithms, AMBO eliminates predefined parameters that may introduce subjectivity from planners. Beyond the algorithm, the proposed model incorporates a noise term to account for inevitable simulation deviations, enabling the identification of better-performing planning results that meet the unique requirements of cold regions. What's more, the characteristics of electric-thermal coupling scenarios are captured and reflected in the operation simulation model to make sure the simulation is close to reality. Numerical simulation verifies the superiority of the proposed approach in generating a more diverse and evenly distributed Pareto front in a sample-efficient manner, providing comprehensive and objective planning choices.
\end{abstract}
\nomenclature[AC]{CSH}{Combination storage heater}


\printnomenclature[2cm]
\nomenclature[AR]{RES}{Renewable energy source}
\nomenclature[AC]{CHP}{Combined heat and power}
\nomenclature[Bt]{$t \in [1,...,T]$}{Index of time interval}
\nomenclature[Bj]{$j \in [1,...,J]$}{Index of iteration time}
\nomenclature[Bk]{$k \in \Omega$}{Index of equipment kind}
\nomenclature[Bi]{$i \in [1,...,N_k]$}{Index of equipment number}
\section{Introduction}
\IEEEPARstart{T}{he} global landscape of renewable energy sources (RESs) has undergone a remarkable transformation over the past few decades. With the urgent need to mitigate climate change and reduce greenhouse gas emissions, countries worldwide have significantly increased their investments in RES, such as wind, solar, and hydropower, including cold regions like northern Europe, North America, and northeast China \cite{olabi2022renewable}. These regions have made substantial progress in integrating RESs into their energy systems by leveraging their abundant natural resources, like strong winds and extended daylight hours during certain seasons, to harness renewable energy effectively.

However, the integration of RESs in regions like northeast China faces significant challenges, particularly due to the prevalence of district heating networks. In these areas, combined heat and power (CHP) generators are the primary source of heat for residential heating due to their high efficiency. However, the operational constraints of CHP systems, which must maintain a constant output to meet the heat load demand, limit their flexibility to adjust power generation \cite{lai2020operation}. This inflexibility leads to a substantial issue of RES power curtailment. As a result, such areas often experience the highest wind curtailment rates over 30\%, turning into an urgent problem to be resolved.


\subsection{Literature Review and Research Gaps}
The power curtailment of RES, driven by the coupling between heat and electricity demand, has garnered attention from researchers \cite{Ding_wind}. Various solutions have been proposed from multiple perspectives, including mechanical improvements to CHP generators \cite{9409901}, enhanced dispatch models \cite{9815530}, fuel cell integration \cite{9826833}, and heat demand response strategies \cite{10075469,yi2021improving}. Among these, the configuration of additional heat sources emerges as a critical factor.

In \cite{9990600}, two types of thermal energy storage (TES) systems are proposed to address multi-scale uncertainties in cold regions, each designed to operate on different time scales.
\citet{xu2020feasibility} propose a hybrid energy system combining a solar air collector, an air-source heat pump, and energy storage for use in cold regions. In \cite{9844159}, an energy storage system incorporating both TES and battery energy storage is installed in buildings, enabling participation in demand response programs.
Current research on heat source planning encompasses a range of equipment on both the generation and consumption sides. To maximize societal welfare, a collaborative capacity planning model is required, which simultaneously considers the planning of these equipment.
\nomenclature[AT]{TES}{Thermal energy storage}

In power system planning, particularly concerning the planning of heat sources, achieving a balance between economic viability and sustainability often requires trade-offs, as accomplishing the two objectives simultaneously is highly challenging. Traditional models typically utilize a weighted sum approach to integrate economic and sustainability objectives—such as investment \cite{10102312,9822978,9920235}, RES power curtailment \cite{10102312}, and carbon taxation \cite{9822978}—into a single objective function. While this method is straightforward, it may overlook complex trade-offs and introduce subjectivity, as planners must determine the weights based on personal judgment.

Conversely, multi-objective optimization (MOO) tackles the complexity of these factors by identifying Pareto-optimal solutions, which offer a range of options instead of a single compromise. MOO empowers planners to make more informed decisions by integrating insights from the Pareto front with their expertise \cite{TIAN2024129612}.
\nomenclature[AM]{MOO}{Multi-objective optimization}

MOO algorithms used in capacity planning problems are predominantly categorized into two types. The first type employs adaptive weights for objectives, exemplified by the adaptive weighted-sum algorithm \cite{AWS} and its enhanced variant \cite{EAWS}. These dynamic weights eliminate the subjectivity of planners but often result in an uneven distribution of Pareto front points and necessitate manual adjustment of certain parameters. The second type is derived from heuristic search methods, such as NSGA-II \cite{NSGAII} and MOAVOA \cite{MOAVOA}.
\textcolor{revise_blue}{
        Both of these types can be integrated with Bayesian optimization to form MOBO algorithms. For instance, in \cite{wang_high-dimensional_2024}, a modified version of PBO  \cite{PBO} is proposed to address high-dimensional optimization problems in transportation systems. Similarly, in  \cite{WANG2023233602}, basic Bayesian optimization is combined with varying weights to identify optimal charging protocols.
        Despite these advancements, the application of MOBO within the power system domain remains largely underexplored. Moreover, the utilization of Bayesian optimization in power systems has primarily focused on hyperparameter tuning for machine learning algorithms \cite{kianpoor_home_2024, jia_convopf-dop_2023, Ding_Bayesian} rather than directly optimizing power system models.
    }

While the heuristic algorithms are effective in handling nonlinear problems, they face challenges in determining the appropriate population size \cite{PBO}, computational efficiency, and dealing with the overlapping points in the Pareto fronts \cite{EAWS}. Importantly, neither category directly addresses the distribution of Pareto fronts, which is crucial for facilitating a more informed decision-making process.

In addition to optimization methods, heat source planning encounters several challenges. Capacity planning models are crucial for preparing power systems to meet future demands more effectively. However, the actual scenarios these systems will encounter are unpredictable, leading to deviations between computer simulations and real-world performance. This issue is particularly pronounced in systems with high penetration of RESs, whose output power is highly variable. Traditional methods often rely on \textit{typical scenarios} \cite{WANG2023121212} or \textit{extended dispatch cycles} \cite{9695353} to mitigate this variability, but these approaches tend to compromise either objectivity or computational efficiency. Furthermore, few studies address the operational scenarios specific to cold regions, where heat load, electric load, and RES output are deeply interdependent. This underscores the necessity for a specially designed time series scenario generation method to effectively tackle these unique challenges.


\subsection{Summary of Major Contributions}
To address the challenges outlined above, this study introduces a collaborative capacity planning approach, underpinned by the development of an adaptive multi-objective Bayesian optimization (AMBO) algorithm. This method effectively produces an informative and objective Pareto front, detailing capacity allocation schemes for electric boilers, TES, heat pumps, and combined storage heaters (residential heaters integrated with small-capacity heat storage). The integration of multiple components from both the generation and consumption sides enhances the potential of hybrid heat sources to increase the accommodation of RES power within the power system while minimizing economic costs.
\nomenclature[AA]{AMBO}{Adaptive multi-objective Bayesian optimization}

In comparison to existing works, three main contributions are presented in the study, summarized as follows:
\textcolor{revise_blue}{
\begin{itemize}
    \item[i)] The study advances basic Bayesian-based MOO algorithms by introducing the AMBO algorithm, applied to capacity planning problems for the hybrid heat sources needed in cold regions. This innovative approach is able to generate a diverse Pareto front in a sample-efficient manner, getting rid of predetermined parameters that could introduce planner bias.
    \item[ii)] A noise-model based approach is proposed to handle the simulation deviation caused by RES variability, ensuring the capacity allocation of heat sources from the generation and consumption sides can adapt to the electric-thermal coupling system and release operation flexibility in the cold regions.
    \item[iii)] The study proposes a time series scenario generation method considering the innate coupling in the heat load, electricity load and RES output. K-medoids clustering is combined with the modification of the key characteristics to make sure the generated scenarios more closely reflect real operational environments.
\end{itemize}
}
\section{Problem Formation}
\subsection{Wind Curtailment in the Cold Areas}
\begin{figure*}[!htb]
    \centering
    \includegraphics[scale=1]{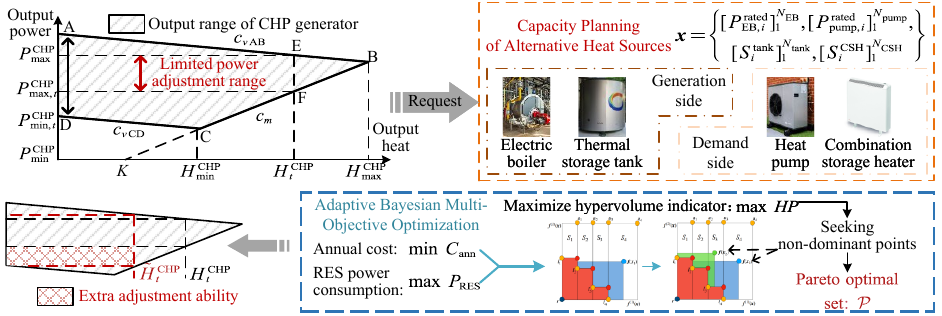}
    \caption{The motivation for planning of the hybrid heat sources.}
    \label{fig:overall}
\end{figure*}

In cold regions, the integration of wind energy into the power grid is significantly hindered by the operational constraints of CHP generators, which are essential for providing residential heating. As illustrated in Fig. \ref{fig:overall}, CHP generators must follow a heat-lead output model. The constant and abundant heat load in cold regions leads to a limited power adjustment range. This inflexibility results in substantial RES power curtailment and highlights the challenge that the need to sustain heat output restricts the ability to establish a low-carbon power system. Consequently, this mismatch between heat demands and RES output power necessitates the use of hybrid heat sources to enhance the power adjustment capabilities of CHP generators through an effective capacity planning approach.

\subsection{Relationship between the Noise and Deviation of Operation Simulation}\label{noise_introduction}

In capacity planning problems, it is necessary to evaluate a certain capacity allocation scheme's performance based on the objectives set by the planner.
Unfortunately, the involvement of RESs makes it tricky to form an accurate evaluation method because the future RES scenarios that confront a power system to be planned hold innate unpredictability and boundlessness.
Such characteristics introduce an unavoidable deviation between the simulation and real-world environment in the evaluation, indicating that the real economic performance of a power system can hardly be precisely evaluated.
Typical planning approaches rely on representative RES scenarios to replace the unpredictable real scenarios. The long term historical RES scenarios or a set of typical scenarios are utilized to alleviate the deviations.
However, while historical RES scenarios or those derived from them provide valuable insights, they cannot perfectly replicate the complexities of the future. Furthermore, numerous other uncertain factors—such as load power fluctuations, heat load variability, and changes in grid parameters—are difficult to simulate accurately.

Taking the economic objective, the annual cost as an example, the simulation deviation is essentially a noise function added to the real objective value, as shown in \eqref{noise}:
\begin{equation}\label{noise}
    C^\mathrm{ob}_\mathrm{ann}(\bm x\mid \bm{S}_\mathrm{typ}) = C_\mathrm{ann}(\bm x) + \epsilon(\bm{x} \mid \bm{S}_\mathrm{typ}),
    \nomenclature[CS]{$\bm{S}_\mathrm{typ}$}{The generation typical scenario containing data of electric load, heat load, and RES output}
\end{equation}
with
\begin{equation}
    \bm{S}_\mathrm{typ} = [P^\mathrm{wind}_{i,t}, P^\mathrm{PV}_{i,t}, P^\mathrm{load}_{t}, H^\mathrm{load}_{t}]
\end{equation}
Equation \eqref{noise} illustrates a scenario in which the typical scenario method is used to evaluate the annual cost of a capacity allocation scheme \(\bm{x}\). The annual cost of \(\bm{x}\) in the real environment, \(C_\mathrm{ann}(\bm{x})\), cannot be accurately determined; instead, only the outcome evaluated under the typical scenario \(\bm{S}_\mathrm{typ}\) can be obtained through simulation. However, this outcome is inevitably affected by the noise term \(\epsilon(\bm{x} \mid \bm{S}_\mathrm{typ})\), as depicted in Fig. \ref{fig:noise_influence}.

\onecolfigure{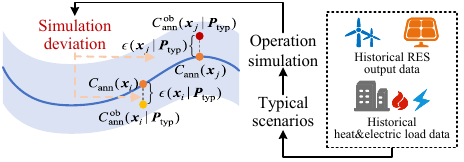}{The influence of simulation deviation on the objective function.}

Whereas the noise term has a detrimental effect of misleading the optimization process, the proposed AMBO algorithm is designed to effectively address this issue, ensuring that the capacity allocation scheme developed by the planning model performs well under real operational conditions.

\subsection{Construction of Time Series Operation Scenarios}\label{sect:scenario_generation}
Chronological RES output and load curves are widely utilized in planning models to assess the performance of capacity allocation schemes. However, a long time range must be considered to ensure reliable results, thereby reducing simulation deviation to an acceptable level. This requirement often leads to unmanageable computational complexity. \textcolor{revise_blue}{Therefore, choosing a limited number of representative operational scenarios from historical data proves to be a more effective and practical method for addressing the deviation.}

\textcolor{revise_blue}{In typical power systems, operation scenarios focus on electricity load, wind, and solar power. However, in cold regions, heat demand is weather-dependent and coupled with RESs. The method proposed in this paper integrates heat load into the scenario generation, considering the interplay between heating demand and RES. This approach reflects the unique challenges of cold regions, where both electricity and heat need to be managed together for efficient system operation.}

In the model proposed in the study, each month in the heating season is represented by a typical day selected by the K-medoids clustering method.
Four crucial characteristics are considered in the clustering process: the variance and mean values of both heat load and net load, the latter being defined as electric load minus RES output.
After selecting the typical day curves for each month, necessary adjustments to the parameters \(a\) and \(b\) in \eqref{typ} are made to ensure that the variance and mean values of the typical curves align with those of the original curves.
\begin{equation}\label{typ}
    \bm{S}_\mathrm{typ} = (\bm{S}'_\mathrm{typ})^a +b.
    \nomenclature[CS]{$\bm S'_\mathrm{typ}$}{The generated scenarios from clustering without adjustment}
\end{equation}
\section{Evaluation Model}\label{sect:evaluation model}
\subsection{Optimization }
The scheme evaluation model comprises two objectives: the economic objective $C_\mathrm{ann}$ and the sustainability objective $P_\mathrm{RES}$. The two objectives are acquired through an operation simulation model introduced under the times series operation scenarios. As formerly mentioned in Section \ref{noise_introduction}, the results directed from the operation simulation model are corrupted by noise because of the simulation deviation.
The detailed formulation of the two objectives is introduced in \eqref{annula_cost}-\eqref{component_of_investment}.
\subsubsection{Annual cost}
\begin{equation}\label{annula_cost}
    \mathop{\arg\min}_{\bm x} \phantom{1} C_\mathrm{ann} = C_\mathrm{inv} + C_\mathrm{gen} ,
    \nomenclature[CC]{$C_\mathrm{ann}$}{Annual cost of a power system}
    \nomenclature[CC]{$C_\mathrm{inv}$}{Equivalent annual investment}
\end{equation}
with
\begin{gather*}
    \bm x = \{[P_{\mathrm{EB},i}^\mathrm{rated}]_1^{N_\mathrm{EB}}, [P_{\mathrm{pump},i}^\mathrm{rated}]_1^{N_\mathrm{pump}}, [S^\mathrm{TES}_{i}]_1^{N_{\mathrm{TES}}}, [S^\mathrm{CSH}_{i}]_1^{N_\mathrm{CSH}}\},
    \nomenclature[CP]{$P_{k,i}^\mathrm{rated}$}{Rated power of \ith equipment $k$}
    \nomenclature[CS]{$S^\mathrm{TES/CSH}_{i}$}{Capcity of \ith TES and CSH}
    \nomenclature[DN]{$N_k$}{Number of equipment $k$}
\end{gather*}
\begin{equation}
    C_\mathrm{inv} = \sum_{k \in \Omega} C_{k}{\mathrm{CPR}}_{k} + C_\mathrm{O\&M},
    \nomenclature[DC]{$\mathrm{CPR}_{k}$}{Capital recovery factor of equipment $k$}
    \nomenclature[CC]{$C_{k}$}{Total investment of equipment $k$}
\end{equation}
\begin{equation}\label{component_of_investment}
    \left\{
    \begin{aligned}
         & C_k = \sum\nolimits_{i}^{N_k} p_k P_{k,i}^\mathrm{rated} \text{ or } C_k = \sum\nolimits_{i}^{N_k} p_k S_k \\
         & C_\mathrm{O\&M}= \sum\nolimits_{i}^{N_k} \kappa_k C_k                                                      \\
         & \mathrm{CPR}_k = \frac{\tau(1+\tau)^{T_{\mathrm{life},k}}}{(1+\tau)^{T_{\mathrm{life},k}}-1}
    \end{aligned}
    \right. k \in \Omega,
    \nomenclature[Dp]{$p_{k}$}{Invement of equipment $k$ per unit}
    \nomenclature[DT]{$T_{\mathrm{life},k}$}{Desigend lifetime of equipment $k$}
    \nomenclature[D]{$\tau$}{Interest rate}
    \nomenclature[D]{$\kappa_k$}{Operation and maintenance cost rate}
    \nomenclature[CC]{$C_\mathrm{O\&M}$}{Operation and maintenance cost}
\end{equation}
\begin{equation*}
    \Omega = \{\mathrm{EB, pump, TES, CSH}\}.
\end{equation*}
\subsubsection{Consumed RES power}
\begin{equation}\label{RES_power}
    \mathop{\arg\min}_{\bm x} \phantom{1} P_\mathrm{RES} = \sum_{t}^{T} \big(\sum_{i}^{N_\mathrm{wind}}P^\mathrm{wind}_{i,t} + \sum_{i}^{N_\mathrm{PV}}P^\mathrm{PV}_{i,t}\big)
    \nomenclature[CP]{$P_\mathrm{RES}$}{Annual consumed RES power}
\end{equation}
\subsection{Operation Simulation Model}
The operation simulation model builds upon the original work in \cite{conference}. Its objective is to achieve a minimum generation cost, including the expenses associated with traditional generators, CHP generators, electric boilers, and CSHs.
\begin{equation}
    \mathop{\arg\min}_{\bm P, \bm{H}} \phantom{1}	C_\mathrm{gen} = C_\mathrm{tra} + C_\mathrm{CHP} + C_\mathrm{EB} + C_\mathrm{CSH},
    \nomenclature[CC]{$C_\mathrm{gen}$}{Annual generation cost}
    \nomenclature[CC]{$C_\mathrm{tra/CHP}$}{Annual generation cost of traditional and CHP generator}
    \nomenclature[CC]{$C_\mathrm{EB/CSH}$}{Annual generation cost of electric boiler and CSH}
\end{equation}
with
\begin{equation*}
    \left\{
    \begin{aligned}
         & \bm P = [P^{\mathrm{tra}}_{i,t},P^\mathrm{wind}_{i,t}, P^\mathrm{PV}_{i,t},P^\mathrm{CHP}_{i,t},P^\mathrm{EB}_{i,t},P^\mathrm{CSH}_{i,t}, P^{\mathrm{pump}}_{i,t}] \\
         & \bm H = [H^\mathrm{CHP}_{i,t},H^\mathrm{pump}_{i,t},H^\mathrm{EB}_{i,t},H^\mathrm{CSH,in/out}_{i,t},H^\mathrm{TES,in/out}_{i,t}]
    \end{aligned}\right. ,
    \nomenclature[CP]{$P_{i, t}^{\mathrm{CHP}}$}{Output power of $i^\mathrm{th}$ CHP generator at time $t$}
    \nomenclature[CP]{$P_{i, t}^{\mathrm{tra}}$}{Output power of $i^\mathrm{th}$ traditional generator at time $t$}
    \nomenclature[CP]{$P_{i, t}^{\mathrm{wind/PV}}$}{Output power of $i^\mathrm{th}$ wind farm and photovoltaic station at time $t$}
    \nomenclature[CP]{$P_{i, t}^{\mathrm{EB}}$}{Consumed power of $i^\mathrm{th}$ electric boiler at time $t$}
    \nomenclature[CP]{$P_{i, t}^{\mathrm{CSH}}$}{Consumed power of $i^\mathrm{th}$ CSH at time $t$}
    \nomenclature[CP]{$ P^{\mathrm{pump}}_{i,t} $}{Consumed power of $i^\mathrm{th}$ pump at time $t$}
    \nomenclature[CH]{$H_{i, t}^{\mathrm{CHP}}$}{Output heat of $i^\mathrm{th}$ CHP generator at time $t$}
    \nomenclature[CH]{$H_{i, t}^{\mathrm{pump}}$}{Output heat of $i^\mathrm{th}$ heat pump at time $t$}
    \nomenclature[CH]{$H_{i, t}^{\mathrm{EB}}$}{Output heat of $i^\mathrm{th}$ electric boiler at time $t$}
    \nomenclature[CH]{$H_{i, t}^{\mathrm{CSH,in/out}}$}{Input and output heat of $i^\mathrm{th}$ CSH at time $t$}
    \nomenclature[CH]{$H_{i, t}^{\mathrm{TES,in/out}}$}{Input and output heat of $i^\mathrm{th}$ TES at time $t$}
\end{equation*}
\begin{equation}
    \left\{
    \begin{aligned}
        C_\mathrm{tra} & = T_\mathrm{d} \sum\nolimits_{i}^{N_{\mathrm{tra}}}\sum\nolimits_{t}^T \big[\mathrm{c}_{1,i}^\mathrm{tra} (P_{i,t}^\mathrm{tra})^{2}     \\
                       & +\mathrm{c}_{2,i}^\mathrm{tra} P_{i,t}^\mathrm{tra} +\mathrm{c}_{3,i}^\mathrm{tra}  \big]                                                \\
        C_\mathrm{CHP} & = T_\mathrm{d} \sum\nolimits_{i}^{N_{\mathrm{CHP}}}\sum\nolimits_{t}^T\big[ c_{1,i}^{\mathrm{CHP}}\left(P_{i, t}^{\mathrm{CHP}}\right)^2 \\
                       & +c_{2,i}^{\mathrm{CHP}} P_{i, t}^{\mathrm{CHP}}+c_{3,i}^{\mathrm{CHP}} +c_{4,i}^{\mathrm{CHP}}\left(H_{i, t}^{\mathrm{CHP}}\right)^2     \\
                       & +c_{5,i}^{\mathrm{CHP}} H_{i, t}^{\mathrm{CHP}} +c_{6,i}^{\mathrm{CHP}} H_{i, t}^{\mathrm{CHP}} P_{i, t}^{\mathrm{CHP}}\big]             \\
        C_\mathrm{EB}  & = c_\mathrm{EB}  T_\mathrm{d}\sum\nolimits_{t}^{T} \sum\nolimits_{i}^{N_\mathrm{EB}} P^{\mathrm{EB}}_{i,t}                               \\
        C_\mathrm{CSH} & =  c_\mathrm{CSH}  T_\mathrm{d}\sum\nolimits_{t}^{T} \sum\nolimits_{i}^{N_\mathrm{CSH}} P^{\mathrm{CSH}}_{i,t}
    \end{aligned}\right. .
\end{equation}

The operation simulation model contains the following constraints detailed in \eqref{power_balance}-\eqref{CSH_initial_state}.
\subsubsection{Power balance constraint}
\begin{equation}\label{power_balance}
    \begin{gathered}
        \sum_i^{N_\mathrm{CHP}} P_{i,t}^\mathrm{CHP}+\sum_i^{N_\mathrm{wind}}P_{i,t}^\mathrm{wind}+\sum_i^{N_\mathrm{PV}}P_{i,t}^\mathrm{PV}+\sum_i^{N_\mathrm{tra}} P_{i,t}^\mathrm{tra}=P^{\mathrm{load}}_t\\+\sum_i^{N_\mathrm{pump}} P^{\mathrm{pump}}_{i,t}
        +\sum_i^{N_\mathrm{EB}} P^{\mathrm{EB}}_{i,t} + \sum_i^{N_\mathrm{TES}}\rho^\mathrm{TES}_i (H^{\text {TES, in }}_{i,t}+ H^{\text {TES, out}}_{i,t})\\
        +\sum_i^{N_\mathrm{CSH}}[\rho^\mathrm{CSH}_i (H^{\text {CSH,in}}_{i,t}+ H^{\text{CSH,out}}_{i,t})+P^\mathrm{CSH}_{i,t} ].
    \end{gathered}
    \nomenclature[CP]{$P^{\mathrm{load}}_t$}{Load power at time $t$}
    \nomenclature[D]{$\rho^\mathrm{CSH}_i$, $\rho^\mathrm{TES}_i$}{Heat to power coefficient of $i^\mathrm{th}$ CSH and TES}
\end{equation}
\subsubsection{Heat balance constraints}
The thermal inertia and heat loss of the district heating network are considered in the heat balance constraints. Equation \eqref{heat_balance}, \eqref{heat_load} denote the input and output constraint for a regional heat network. Equation \eqref{heat_inertia} captures the thermal storage and inertia constraints, and the upper and lower bound heat energy constraint is defined in \eqref{heat_bound} for the heat network to make sure the comfort of residents is not compromised.
\begin{equation}\label{heat_balance}
    \begin{gathered}
        H^{\text{net,in}}_t=\sum_i^{N_\mathrm{CHP}}H^{\mathrm{CHP}}_{i,t}+\sum_i^{N_\mathrm{pump}}H^{\text{pump}}_{i,t}\\
        + \sum_i^{N_\mathrm{EB}}H^{\mathrm{EB}}_{i,t}+\sum_i^{N_\mathrm{TES}}H^{\text {TES,out}}_{i,t} + \sum_i^{N_\mathrm{CSH}}H^{\text {CSH,out}}_{i,t},
    \end{gathered}
    \nomenclature[CH]{$H^{\text{net,in/out}}_t$}{Input and output heat of district heating netwrok at time $t$}
\end{equation}
\begin{equation}\label{heat_load}
    H^{\text{net,out}}_t=H^{\mathrm{load}}_t,
    \nomenclature[CH]{$H^{\mathrm{load}}_t$}{Heat load at time $t$}
\end{equation}
\begin{equation}\label{heat_inertia}
    E^{\text{net}}_t=E^{\text{net}}_{t-1}+(1-\lambda_\mathrm{net}) H^{\text{net,in}}_t-H^{\mathrm{net,out}}_{t+T_{\text {delay}}} ,
    \nomenclature[D]{$\lambda_{\text{net}}$}{Heat loss coefficient of heat network}
    \nomenclature[DT]{$T_{\text{delay}}$}{Time delay in the heat transfer process}
\end{equation}
\begin{equation}\label{heat_bound}
    E^{\text{net}}_{\min} \le E^{\text{net}}_t \le 	E^{\text{net}}_{\max} .
    \nomenclature[DE]{$E^{\text{net}}_{\min/\max}$}{Upper and lower limit of heat energy of district heat network}
    \nomenclature[CE]{$E^{\text{net}}_t$}{The energy stored in the heat network at time $t$}
\end{equation}
\subsubsection{Traditional generator constraints}
\begin{equation}
    P_{\min,i}^\mathrm{tra} \le P_{i,t}^\mathrm{tra} \le P_{\max,i}^\mathrm{tra} ,
\end{equation}
\begin{equation}
    -P_{\mathrm{ramp},i}^\mathrm{tra} \le P^\mathrm{tra}_{i,t} - P^\mathrm{tra}_{i,t-1} \le P_{\mathrm{ramp},i}^\mathrm{tra}.
    \nomenclature[DP]{$P_{\min/\max,i}^{k}$}{Upper and lower limit on the output power of $i^\mathrm{th}$ equipment $k$}
    \nomenclature[DP]{$P_{\mathrm{ramp},i}^\mathrm{tra/CHP}$}{Ramping rate limitation of $i^\mathrm{th}$ traditional generator and CHP generator}
\end{equation}
\subsubsection{Combined heat power generator constraints}
The CHP generator's operational characteristics hold a distinct coupling between its electrical and thermal outputs as shown in Fig. \ref{fig:overall}, which are formulated in \eqref{CHP_range1}-\eqref{CHP_range3}.
\begin{equation}\label{CHP_range1}
    \begin{gathered}
        \max\{P^{\mathrm{CHP}}_{{\min},i} - c^{vCD}_i H^\mathrm{CHP}_{i,t}, c^m_i H^\mathrm{CHP}_{i,t} + c^\mathrm{k}_i\}  \\ \le P^\mathrm{CHP}_{i,t} \le P^\mathrm{CHP}_{i,t} - c^{cAB}_i H^\mathrm{CHP}_{i,t}
        \nomenclature[Dc]{$c^{vCD/m/cAB/k}_i$}{Feasible operating region parameters of $i^\mathrm{th}$ CHP generator}
    \end{gathered},
\end{equation}
\begin{equation}\label{CHP_range2}
    P_{\mathrm{min},i}^\mathrm{CHP} \le P^\mathrm{CHP}_{i,t} \le P_{\mathrm{max},i}^\mathrm{CHP},
\end{equation}
\begin{equation}\label{CHP_range3}
    H_{\mathrm{min},i}^\mathrm{CHP} \le H^\mathrm{CHP}_{i,t} \le H_{\mathrm{max},i}^\mathrm{CHP},
    \nomenclature[DH]{$H^{k}_{{\min/\max},i}$}{Upper and lower limit on the output heat of $i^\mathrm{th}$ equimpment $k$}
\end{equation}
\begin{equation}
    -P_{\mathrm{ramp},i}^\mathrm{CHP} \le P^\mathrm{CHP}_{i,t} - P^\mathrm{CHP}_{i,t-1} \le P_{\mathrm{ramp},i}^\mathrm{CHP}.
\end{equation}
\subsubsection{Renewable energy source constraints}
In \eqref{wind_power} and \eqref{solar_power}, the $P^\mathrm{wind}_{\max,i,t}$ and $P^\mathrm{PV}_{\max,i,t}$ represent power curves derived from the selected time series typical scenarios. The inequalities indicate that the RES power curtailment is permitted in the operation simulation.
\begin{equation}\label{wind_power}
    0\le P^\mathrm{wind}_{i,t}\le P^\mathrm{wind}_{\max,i,t},
    \nomenclature[CP]{$P^\mathrm{wind/PV}_{\max,i,t}$}{Maximum possible output power of $i^\mathrm{th}$ wind farm and photovoltaic station at time $t$}
\end{equation}
\begin{equation}\label{solar_power}
    0\le P^\mathrm{PV}_{i,t}\le P^\mathrm{PV}_{\max,i,t}.
\end{equation}
\subsubsection{Heat pump constraints}
The water-source heat considered in the study is an efficient equipment that utilizes electricity to gather heat in the air, as formulated in \eqref{pump_heat}.
\begin{equation}\label{pump_heat}
    H^\mathrm{pump}_{i,t} = \mathrm{COP}_i P^\mathrm{pump}_{i,t},
    \nomenclature[DC]{$\mathrm{COP}_i$}{Coefficient of performance of $i^\mathrm{th}$ heat pump}
\end{equation}
\begin{equation}
    P^\mathrm{pump}_{\min,i} \le P^\mathrm{pump}_{i,t} \le P^\mathrm{pump}_{\max,i}.
\end{equation}
\subsubsection{Electric boiler constraints}
Electric boiler is a widely adopted complementary heat source for the CHP generator, it directly transfers electricity into heat as in \eqref{EB_heat}.
\begin{equation}
    P^\mathrm{EB}_{\min,i} \le P^\mathrm{EB}_{i,t} \le P^\mathrm{EB}_{\max,i},
\end{equation}
\begin{equation}\label{EB_heat}
    H^\mathrm{EB}_{i,t} = \beta^{\mathrm{EB}}_i P^{\mathrm{EB}}_{i,t}.
    \nomenclature[D]{$\beta^{\mathrm{EB}}_i$, $\beta^{\mathrm{CSH}}_i$}{Conversion coefficient of $i^\mathrm{th}$ electric boiler and CSH}
\end{equation}
\subsubsection{Thermal energy storage constraints}
The TES considered in this study is widely utilized due to its low cost compared to other forms of energy storage, such as batteries. The energy state, heat input, and output constraints are defined in \eqref{S_tank} to \eqref{initial_state_tank}, accounting for self-heat loss.
\begin{equation}\label{S_tank}
    0 \le Q^\mathrm{TES}_{i,t} \le S^\mathrm{TES}_i,
\end{equation}
\begin{equation}
    -H^\mathrm{TES,in}_\mathrm{max} \le (1-\eta^\mathrm{TES}_i) Q^\mathrm{TES}_{i,t-1} - Q^\mathrm{TES}_{i,t} \le H^\mathrm{TES,out}_\mathrm{max} ,
    \nomenclature[DH]{$H^\mathrm{TES,in/out}_\mathrm{max}$}{Upper limit on the input and output heat of $i\mathrm{th}$ TES system}
    \nomenclature[CQ]{$Q^\mathrm{TES/CSH}_{i,t}$}{Stored energy of $i^\mathrm{th}$ TES system and CSH at time $t$}
    \nomenclature[D]{$\eta^\mathrm{TES}_i$, $\eta^\mathrm{CSH}_i$}{Self-discharge rate of $i^\mathrm{th}$ TES and CSH}
\end{equation}
\begin{equation}\label{initial_state_tank}
    Q^\mathrm{TES}_0 = Q^\mathrm{TES}_T.
\end{equation}
\subsubsection{Combination storage heater constraints}
The CSH operates like an electric boiler combined with a heat storage tank, detailed characteristics are demonstrated below:
\begin{equation}
    H^\mathrm{CSH, in}_{i,t} = \beta^\mathrm{CSH}_i P^\mathrm{CSH}_{i,t},
\end{equation}
\begin{equation}
    -H^\mathrm{CSH,in}_\mathrm{max} \le (1-\eta^\mathrm{CSH}_i) Q^\mathrm{CSH}_{i,t-1} - Q^\mathrm{CSH}_{i,t} \le H^\mathrm{CSH,out}_\mathrm{max},
\end{equation}
\begin{equation}\label{CSH_initial_state}
    Q^\mathrm{CSH}_0 = Q^\mathrm{CSH}_T.
\end{equation}
\section{Proposed Adaptive Multi-objective Bayesian Optimization Algorithm}

As formerly introduced in  Section \ref{noise_introduction}, the outcomes from the operation simulation based on the typical scenarios are corrupted by the noise term due to the simulation deviation. In this section, a detailed mathematical process will demonstrate how the proposed AMBO algorithm carries out an efficient optimization under such noisy circumstance.

As in \cite{noisyMOBO}, the AMBO algorithm gains an edge compared with traditional MOO algorithms in that it directly optimizes the distribution of the Pareto fronts through the maximization of the hypervolume indicator, a widely used measure for the assessment of the quality of Pareto fronts.

At the beginning of each iteration, a capacity allocation scheme generated by the AMBO algorithm is evaluated in the evaluation model, yielding two objective values.
Next, the Gaussian process, a non-parametric model, is applied to infer a distribution over functions that best describes the relationship between capacities and each objective function based on all the acquired capacity allocation schemes and their corresponding objective values.

The distributions provide insights into the latent optimal point, which can be identified by maximizing the function known as the \textit{expected hypervolume improvement}.
Although the distributions generated by the Gaussian process are subject to unavoidable errors, the iterative process of inference and evaluation yields cumulative outcomes that help them converge toward the actual objective functions.

The detailed optimization process of AMBO is introduced in the following subsections.
\subsection{Initialization}

To initiate the process, a set of $N_0$ capacity allocation schemes is randomly selected and evaluated, bringing an initial dataset $\bm D_0$ containing the observed objective function values $\bm Y_0$ and the corresponding scheme $\bm X_0$.
Specific definitions of $\bm{X}_0 $ and $ \bm Y_0$ are provided as follows:
\begin{equation}\label{D}
    \bm D_0 = \{\bm X_0 ,\bm Y_0\} ,
    \nomenclature[CD]{$\bm D_j$}{Acquired dataset at $j^\mathrm{th}$ iteration}
    \nomenclature[CX]{$\bm X_j$}{Evaluated scheme set at $j^\mathrm{th}$ iteration}
    \nomenclature[CY]{$\bm Y_j$}{Observed objective values at $j^\mathrm{th}$ iteration}
\end{equation}
where
\begin{gather}
    \bm X_0 = \{\bm x_i\}_1^{N_0} = \{ [P_{\mathrm{EB},i}^\mathrm{rated}]_1^{N_\mathrm{EB}}, [P_{\mathrm{pump},i}^\mathrm{rated}]_1^{N_\mathrm{pump}}, \notag \\
    [S^\mathrm{TES}_{i}]_1^{N_{\mathrm{TES}}}, [S^\mathrm{CSH}_{i}]_1^{N_\mathrm{CSH}}  \}_1^{N_0},\label{initalx}
\end{gather}
\begin{gather}
    \bm Y_0 =\{\bm y_i\}_1^{N_0}= \{ [f_1(\bm x_i) + \epsilon_{1},f_2(\bm x_i)+ \epsilon_{2}] \} \notag \\
    = \{[C^\mathrm{ob}_\mathrm{ann}(\bm x_i), -P_\mathrm{RES}^\mathrm{ob}(\bm x_i) ]\}_1^{N_0} \label{initaly}.
\end{gather}

\textcolor{revise_blue}{Equations \eqref{initalx} and \eqref{initaly} demonstrate the connection between the proposed algorithm and the evaluation model detailed in the preceding section. The vector $\bm{X}$ comprises sets of equipment capacity, where each set represents a capacity allocation scheme. Conversely, $\bm{Y}$ contains the values of two objectives, as defined in \eqref{annula_cost} and \eqref{RES_power}, corresponding to each scheme represented in $\bm{X}$. The primary objective for optimization within the AMBO algorithm is the hypervolume indicator derived from the elements of $\bm{Y}$, which will be elucidated in Algorithm \ref{alg:xcand}.
}

Once the outcome set is initially acquired or updated, the posterior distributions in the $j$ iteration $\bm p_J= [p_{j,i}(f_i\mid \bm D_j, \sigma_{\mathrm{n},i})]_1^2$ are created using the Gaussian process over each objective function $f_i$.
\begin{equation}\label{GP1}
    f_i(\bm x) \sim p_{j,i}( f_i \mid \bm D_j, \sigma_{\mathrm{n},i}) = \mathcal{N}( \mu _{\bm D_j,i}, \bm \Sigma_{\bm D_j,i}),\phantom{1} i=1,2 ,
\end{equation}
with
\begin{equation}\label{GP2}
    \left\{
    \begin{aligned}
         & \mu _{\bm D_j,i}(\bm x) =  \bm K(\bm x, \bm X_j)[\bm K(\bm X_j, \bm X_j) + \sigma_{\mathrm{n},i} \mathbf I]^{-1} \bm{\hat y}_i                                       \\
         & \bm \Sigma_{\bm D_j,i}(\bm x,\bm x^{'}) =\bm K(\bm x, \bm x^{'})                                                                                                     \\
         & -\bm K(\bm x,\bm X_j)[\bm K(\bm X_j, \bm X_j) + \sigma_\mathrm{n} \mathbf I]^{-1} +\bm K(\bm X_j,\bm x^{'})                                                          \\
         & \bm K(\bm X_j, \bm  X_j)=[k(\bm x_p, \bm x_q)]_{p,q=1}^n                                                                                                             \\
         & = [\sigma^2 \frac{2^{1-v}}{\Gamma(v)}(\frac{\sqrt{2 v}\left\|\bm x_p-\bm x_q\right\|}{\ell})^v K_v(\frac{\sqrt{2 v}\left\|\bm x_p-\bm x_q\right\|}{\ell})]_{p,q=1}^n \\
    \end{aligned}
    \right. ,
    \nomenclature[CN]{$\mathcal{N}( \mu, \bm \Sigma)$}{Multivariate normal distribution with mean vector $\mu$ and covariance matrix $\bm \Sigma$}
    \nomenclature[C]{$\mu _{\bm D,i}$}{Mean vector of a multivariate normal distribution}
    \nomenclature[C]{$ \bm \Sigma_{\bm D,i}$}{Covariance matrix of a multivariate normal distribution}
    \nomenclature[DI]{$\mathbf{I}$}{Identity matrix}
    \nomenclature[Cy]{$\bm{\hat y}_i$}{Normalized version of $\bm{ y}_i$}
    \nomenclature[Dk]{$k(\bm x_i, \bm x_j)$}{The Matérn kernel function with three kernel parameters $\sigma$, $v$ and $ l$}
    \nomenclature[D]{$\sigma$, $v$, $l$}{Kernel parameters of the Matérn kernel function}
    \nomenclature[D]{$\Gamma$}{The gamma function}
    \nomenclature[DK]{$K_v$}{The modified Bessel function of the second kind}
\end{equation}
{\noindent where the three kernel parameters $\sigma$, $v$ and $ l$ are determined by maximum likelihood estimation.}
\subsection{Seeking of the Candidate Point}
The generated posterior distributions are used to maximize the the acquisition function known as \textit{noisy expected hypervolume improvement}, determining the candidate capacity allocation scheme $\bm x_\mathrm{cand}$ to be evaluated. The detailed process is shown in Algorithm \ref{alg:xcand}.
\begin{algorithm}
    \caption{Locating the candidate to be evaluated}
    \label{alg:xcand}
    \small
    \begin{algorithmic}[1]
        \Require $\bm r$: reference point; $\bm p$: posterior distributions; $ \bm D= \{[\bm x_i, \bm y_i]_1^n\}$ : evaluated points.
        \Ensure $\bm x_\mathrm{cand}$: candidate point.
        \State \textbf{def} Pareto optimal set: $\mathcal{P}$
        \State \hspace{0.5cm} $\mathcal{P} = \{\bm g(\bm x) \text{ s.t.} \text{ } \nexists \text{ } \bm x'\in \bm X: \bm g(\bm x') <\bm g(\bm x) \}$
        \State \textbf{def} Hypervolume indicator: $HP(\mathcal{P}\mid \boldsymbol{r})$
        \State \hspace{0.5cm} $\boldsymbol{r} \in \mathbb{R}^M: HP(\mathcal{P} \mid \label{HP} \boldsymbol{r})=\lambda_M\left(\bigcup_{\boldsymbol{v} \in \mathcal{P}}[\boldsymbol{r}, \boldsymbol{v}]\right)$ \Comment{$[\boldsymbol{r}, \boldsymbol{v}]$ denotes the hyper-rectangle bounded by vertices}
        \State \textbf{def} Hypervolume improvement: $HVI(\mathcal{P}' \mid \mathcal{P}, r)$
        \State \hspace{0.5cm} $HVI(\mathcal{P}' \mid \mathcal{P}, \boldsymbol{r}) = HP(\mathcal{P}' \cup \mathcal{P} \mid \boldsymbol{r}) - HP(\mathcal{P} \mid \boldsymbol{r})$
        \Procedure{$\max \text{ } HVI$}{$\bm x_\mathrm{cand}$}
        \State $\tilde{\boldsymbol{f}}_t \sim \bm p\left(\boldsymbol{f} \mid \bm D,\bm \sigma_{\mathrm{n}}\right), \text { for } t=1, \ldots N_s$ \Comment{sample from $\bm p$}
        \State $\mathcal{P}_t \leftarrow \{\tilde{\boldsymbol{f}}_t(\bm x) \mid \bm x \in \bm{X}_n, \tilde{\boldsymbol{f}}_t(\bm x)<\tilde{\boldsymbol{f}}_t(\bm x')\text{ } \forall x' \in \bm X_n\}$
        \State $\hat{\alpha}_{\mathrm{NEHVI}}(\boldsymbol{x}) \leftarrow \frac{1}{N} \sum_{t=1}^N HVI \left(\tilde{\boldsymbol{f}}_t(\boldsymbol{x}) \mid \mathcal{P}_t\right)$ \Comment{Approximate ${\alpha}_{\mathrm{NEHVI}}$}
        \State $\bm x_\mathrm{cand} \leftarrow \max \text{ } \hat{\alpha}_{\mathrm{NEHVI}}(\bm x)$ \label{max_NEHVI}
        \EndProcedure
        \Statex \Comment{The detailed calculation of line \ref{HP} and line \ref{max_NEHVI} can be found in \cite{noisyMOBO,EHVI}.}
        \Statex \Comment{\textcolor{revise_blue}{The detailed definitions of variables: $\bm D$ in \eqref{D}, $\bm p$ in \eqref{GP1}, $\bm{X}$ in \eqref{initalx}, $\bm r$ in \eqref{refpoint}.}}
    \end{algorithmic}
\end{algorithm}
Compared with other Bayesian-based MOOs, the repeated sampling process in Algorithms \ref{alg:xcand} from posterior distributions in \eqref{GP1} assists in the neutralization of the simulation deviation. That is how the AMBO algorithm avoids the misleading from the noise term and locates the non-dominant points to form an informative Pareto front.
\subsection{Adaptive Determination of the Crucial Parameters}
The proposed AMBO algorithm is improved from the work in \cite{noisyMOBO} with improvements focusing on the determination of two crucial parameters for optimization, aimed at enhancing the algorithm's robustness and applicability.
\subsubsection{Reference point}
A reference point is utilized in the calculation of the hypervolume indicator, the key metric to be optimized within the algorithm. While there is no universal principle for determining the optimal reference point, it is generally considered appropriate to select a reference point that is marginally inferior to the worst-case scenario to evaluate the quality of the Pareto front.
In the original work from \cite{noisyMOBO}, the reference point is predetermined by the decision-maker. However, considering the complexities of the hybrid heat sources capacity planning model, it costs a huge amount of computational resources to determine the worst case of each objective and limits the algorithm's applicability to planning problems with different models.
As a result, the study establishes the reference point $\bm r$ as an adaptive vector as detailed in \eqref{refpoint}.
\begin{equation}\label{refpoint}
    \bm r = \bm{\hat y}_{\max} - \bm{\hat y}_{\min} * 10\%,
\end{equation}
{\noindent for the minimization problem in the study, the $\bm{\hat y}^{\max}$ stands for the worst-case that has been observed, and the involvement of $\bm{\hat y}^{\min}$ take the range of different objectives into consideration.}
\nomenclature[Cy]{$\bm{\hat y}_{\min/\max}$}{Minimum and maximum values of $\bm{\hat y}$}
\subsubsection{Noise standard deviation}
Noise standard deviation $\bm \sigma_{\mathrm{n}} = [\sigma_{\mathrm{n},1}, \sigma_{\mathrm{n},2}]$ is another parameter that significantly influences the optimization, which is critical for the precision of the probabilistic surrogate model. Similar to the reference point, it is a predetermined vector in \cite{noisyMOBO}.
The noise standard deviation serves as a measure of the noise level in the objective functions. Experimental observations indicate that, in the context of power system planning problems, this standard deviation is influenced by various factors, including the penetration of RESs, load levels, and the system's adjustment capabilities.
Such intricate coupling makes the accurate predetermination of the $\bm \sigma_{\mathrm{n}}$ an impossible task.
To settle the problem, a marginal log likelihood based method is incorporated in the AMBO algorithm, making $\bm \sigma_{\mathrm{n}}$ able to self-update in each iteration as demonstrated in Algorithm \ref{alg:noise_update}.

\begin{algorithm}
    \caption{Estimate noise standard deviation in the Gaussian process}
    \label{alg:noise_update}
    \small
    \begin{algorithmic}[1]
        \Require $\bm{x}$, $\bm{y}$: training data; $n$: number of data points; $k(\bm x_i, \bm x_j)$: kernel function; $\upsilon$: tolerance; $\alpha$: learning rate.
        \Ensure $\bm \sigma_\mathrm{n}$: estimated noise standard deviation.
        \While{$m < M$} \Comment{$M$: number of objectives}
        \State $\sigma_{\mathrm{n},m}\leftarrow \text{random()}>0$ \Comment{initialize with a positive number}
        \Repeat
        \State $\bm K \leftarrow [k(\bm x_i, \bm x_j)]_{i,j=1}^n$
        \State $\bm {K}_{\text{n}} \leftarrow \bm {K} + \sigma^2 \mathbf{I}$
        \State $\mathrm{MLL} \leftarrow -\frac{1}{2} {\bm y_m}^T \bm {K}_{\text{n}}^{-1} \bm{y}_m - \frac{1}{2} \log |\bm {K}_{\text{n}}| - \frac{n}{2} \log(2\pi)$
        \State $\frac{\partial \mathrm{MLL}}{\partial \sigma_{\mathrm{n},m}^2} \leftarrow \frac{1}{2} \operatorname{trace}\left(\bm K_\mathrm{n}^{-1}-\bm K_\mathrm{n}^{-1} \bm{y}_m \bm{y}_m^T \bm K_\mathrm{n}^{-1}\right)$
        \State $\sigma_{\mathrm{n},m} \leftarrow \sqrt{\sigma_\mathrm{\mathrm{n},m}^2+\alpha \frac{\partial \mathrm{MLL}}{\partial \sigma_\mathrm{n}^2}}$\Comment{update by gradient ascent}
        \Until{{$\| \nabla_{\sigma_\mathrm{n}^2} \text{MLL} \| < \upsilon$ \textbf{or} $|\mathrm{MLL}_{\mathrm{new}} - \mathrm{MLL}_{\mathrm{old}}| < \upsilon$}}
        \EndWhile
        \State $\bm \sigma_\mathrm{n}$ $\leftarrow$ $[\sigma_{\mathrm{n},m}]_1^M$
        \State \textbf{return} $\bm \sigma_\mathrm{n}$
        \Statex \Comment{\textcolor{revise_blue}{The detailed definitions of variables: $\bm{y}$ in \eqref{initaly}, $\bm K$ in \eqref{GP2}.}}
    \end{algorithmic}
\end{algorithm}
\subsection{Overall Optimization Process}
\begin{figure}[!htbp]
    \centering
    \includegraphics[scale = 1]{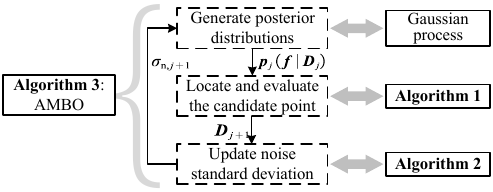}
    \caption{\textcolor{revise_blue}{Relationship of the algorithms presented in the pseudocode}}
    \label{fig:algorithm_flow_chart}
\end{figure}

\textcolor{revise_blue}{The complete optimization process is illustrated in Algorithm \ref{alg:AMBO}, and the relationship of algorithms presented in the pseudocode is presented in Fig. \ref{fig:algorithm_flow_chart}.}
It is important to note that in generating the final Pareto front, the posterior mean \(\bm{\mu}_{\bm{D}_J}\) is used to filter out non-dominant points instead of the objective function values \(\bm{Y}_J\), as the latter are susceptible to noise corruption. The posterior mean provides a more accurate measure for evaluating the performance of a given capacity allocation scheme, as it incorporates latent information from all the acquired evaluation outcomes. The case study in the next section demonstrates the advantages of this approach.
\begin{algorithm}
    \caption{Adaptive Multi-Objective Bayesian Optimization}
    \label{alg:AMBO}
    \small
    \begin{algorithmic}[1]
        \Statex \textbf{Function} AMBO
        \State $j \leftarrow 0$, $\bm D_0 \leftarrow \{\bm X_0, \bm Y_0\}$ \Comment{initialize}
        \While{$j < J$} \Comment{$J$: maximum number of iterations}
        \State GP(${\bm \sigma}_{\mathrm{n},j}$) \textbf{do}
        \State \hspace{0.25cm} $\bm p_j( \bm f \mid \bm D_j )\leftarrow \mathcal{N}(\bm \mu _{\bm D_j}, \bm \Sigma_{\bm D_j})$ \Comment{generate distributions}
        \State $\bm r = \bm{\hat y}^{\max} - \bm{\hat y}^{\min} * 10\%$ \Comment{update reference point}
        \State Algorithm \ref{alg:xcand} \textbf{do}
        \State \hspace{0.3cm} $\bm x_\mathrm{cand} \leftarrow \max \text{ } \hat{\alpha}_{\mathrm{NEHVI}}(\bm x)$ \Comment{find the candidate}
        \State $\bm y_{\mathrm{cand}} \leftarrow \bm f (\bm x_\mathrm{cand}) + \epsilon$ \Comment{evaluate the candidate}
        \State $\bm D_{j+1} \leftarrow \bm D_j \cup \{\bm x_\mathrm{cand}, \bm y_{\mathrm{cand}}\}$
        \Comment{update observation set}
        \State Algorithm \ref{alg:noise_update} \textbf{do}
        \State \hspace{0.3cm} $\bm{\sigma}_{\mathrm{n},j+1} \leftarrow \text{Algorithm 2}(\bm{\sigma}_{\mathrm{n},j})$ \Comment{update noise std}
        \State $j \leftarrow j+1$
        \EndWhile
        \State GP(${\bm \sigma}_{\mathrm{n},J}$) \textbf{do}
        \State \hspace{0.25cm} $\bm p_{J}( \bm f \mid \bm D_{J} )\leftarrow \mathcal{N}(\bm \mu _{\bm D_J}, \bm \Sigma_{\bm D_J})$
        \State $\mathcal{P}^* \leftarrow \{\bm \mu _{\bm D_J}(\bm x^*) \text{ s.t.} \text{ } \nexists \text{ } \bm x \in \bm X_{J}: \bm \mu _{\bm D_J}(\bm x) < \bm \mu _{\bm D_J}(\bm x^*) \}$
        \State \Return $\mathcal{P}^*$
        \Statex \Comment{\textcolor{revise_blue}{The detailed definitions of variables: $\bm D$ in \eqref{D}, $\bm{y}$ in \eqref{initaly}, $\bm p$ in \eqref{GP1}.}}
    \end{algorithmic}
\end{algorithm}
\section{Case Study}
The case study comprises three subsections. First, the generated time series scenarios and the simulation settings are presented. In the second subsection, the advantages of the AMBO algorithm under noiseless conditions are demonstrated in comparison to other MOO algorithms. Finally, the benefits of modeling the simulation deviation as a noise term are emphasized.

\subsection{Simulation and scenario settings}
\subsubsection{Parameters}
\textcolor{revise_blue}{
The simulation model includes two CHP generators and one traditional generator, reflecting the need for combined heat and power systems to meet both heat and electricity demands. To enhance the system's flexibility and accommodate the unique challenges of these regions, the capacity planning incorporates multiple alternative heat sources, including one TES system, one electric boiler, one heat pump, and one CSH. These components are carefully selected to improve the adjustment capability of the system, particularly under severe weather conditions and varying heat loads. Parameters that are closely associated with the planning outcomes are presented in Table \ref{tab:parameter}.
}
\begin{table}[htbp]
    \footnotesize
    \centering
    \caption{Key Parameters of the Capacity Planning Model}
    \begin{threeparttable}
        \begin{tabular*}{0.49\textwidth}{c@{\extracolsep{\fill}}ll}
            \toprule
            Equipment   & Parameter & Value  \\
            \midrule
            \multirow{4}{*}{$1^\mathrm{th}$ CHP generator } &  Fuel cost   & $[c^\mathrm{CHP}_{p,1}]_{p=1}^6= [1.03, 32.74,$  \\[1.5pt]
            &coefficients  &$ 14.62, 0.58, 22.56, 0.15]$ \\[1.5pt]
            & Output range &$[c^{vCD}_1,c^m_1,c^{cAB}_1]$ \\ [1.5pt]
            & coefficients &$=[0.045,0.75,0.15]$ \\ [3pt]
            \multirow{4}{*}{$2^\mathrm{th}$ CHP generator }  & Fuel cost   & $[c^\mathrm{CHP}_{p,2}]_{p=1}^6= [1.09, 38.80,$  \\[1.5pt]
            &coefficients &$ 18.82,  0.61, 24.10, 0.16]$ \\[1.5pt]
            & Output range &$[c^{vCD}_2,c^m_2,c^{cAB}_2]$ \\ [1.5pt]
            & coefficients &$=[0.03,0.72,0.2]$ \\ [3pt]
            Traditional                                     &Fuel cost &  $[c^\mathrm{CHP}_{p,2}]_{p=1}^3 = $ \\ [1.5pt]
            generator                                       & coefficients&  $ [2.44, 35.64, 11.54]  $ \\ [3pt]
            \multirow{4}*{Thermal energy storage}             & Type      & Water-based storage tank  \\[1.5pt]
            & Unit cost & $c_\mathrm{TES}=100000$ \$/MWh  \\[1.5pt]
            & Lifetime &  $ T_\mathrm{TES} = 25 $ year \\[1.5pt]
            & Efficiency & {$\eta^\mathrm{TES}_1 =90\% $} \\[1.5pt]
            \multirow{3}*{Electric boiler}                  & Unit cost &  $c_\mathrm{EB} =  300000$ \$/MW \\[1.5pt]
            & Efficiency & {$\eta^\mathrm{EB}_1 =95\% $} \\[1.5pt]
            & Lifetime  &  $ T_\mathrm{EB} = 25 $ year   \\[3pt]
            \multirow{3}*{Heat pump}                        & Unit cost &  $c_\mathrm{pump} = 3000000 $ \$/MW \\[1.5pt]
            & Efficiency &  $\mathrm{COP}_1 =4 $  \\[1.5pt]
            & Lifetime &   $ T_\mathrm{pump} = 15 $ year   \\[3pt]
            \multirow{2}*{CSH}                              & Unit cost &   $c_\mathrm{CSH} = 50000 $ \$/MW \\[2pt]
            & Lifetime &  $ T_\mathrm{CSH} = 15 $ year    \\
            \bottomrule
        \end{tabular*}
    \end{threeparttable}
    \label{tab:parameter}
\end{table}
\subsubsection{Scenarios}
\begin{figure}[!htbp]
    \centering
    \begin{subfigure}{0.24\textwidth}
        \includegraphics[scale=1]{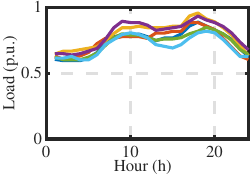}
        \caption{Typical electric load}
        \label{fig:load_curve}
    \end{subfigure}
    \begin{subfigure}{0.24\textwidth}
        \includegraphics[scale=1]{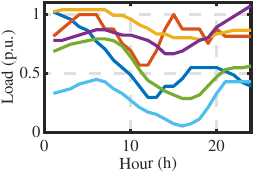}
        \caption{Typical heat load}
        \label{fig:heat_load_curve}
    \end{subfigure}
    \begin{subfigure}{0.24\textwidth}
        \includegraphics[scale=1]{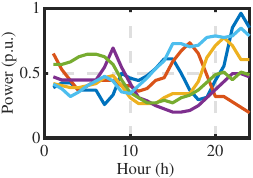}
        \caption{Typical wind power}
        \label{fig:wind_curve}
    \end{subfigure}
    \begin{subfigure}{0.24\textwidth}
        \includegraphics[scale=1]{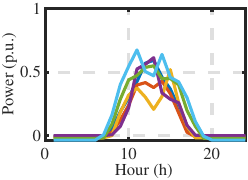}
        \caption{Typical solar power}
        \label{fig:PV_curvez}
    \end{subfigure}
    \caption{The typical curves of electric load, heat load, wind power, and solar power.}
    \label{fig:typical_curves}
\end{figure}

The data used to generate the time series typical scenarios consists of recorded wind and solar output power, along with the corresponding electric and heat loads, from an entire heating season in a province in northeast China. The generated time series scenarios are illustrated in Fig. \ref{fig:typical_curves}.
\subsubsection{Algorithms}
Three additional MOO algorithms are selected for comparison: the original version of AMBO from \cite{noisyMOBO}, PBO from \cite{PBO} (a Bayesian-based MOO that relies on adaptive coefficients), and NSGA-II with a population size of 12.
\textcolor{revise_blue}{\subsubsection{Simulation settings}The simulations were conducted on a computer equipped with a 2.9 GHz Intel Core i9-12900H processor and 32 GB of RAM. The sub-problems in the scheme evaluation model were solved using Gurobi V.10.0. The planning approach was executed in Python 3.10 on a Windows 11 operating system, utilizing the Botorch \cite{NEURIPS2020_f5b1b89d} and Gurobipy packages.}

\subsection{Effectiveness of the Proposed Approach}
In this subsection, only the outcomes directly derived from the operational simulation based on the time series typical scenarios are used to form the Pareto fronts and calculate the hypervolume indicator. This approach allows for a more straightforward comparison of the MOO algorithms.
\begin{table}[!htbp]
    \footnotesize
    \centering
    \caption{\textcolor{revise_blue}{Typical days of each month selected by the scenario generation methods}}
    \begin{threeparttable}
        \begin{tabular*}{0.49\textwidth}{c@{\extracolsep{\fill}}ccc}
            \toprule
              \textcolor{revise_blue}{Month}   & \textcolor{revise_blue}{Proposed} & \textcolor{revise_blue}{\cite{scenario_compare}}*  & \textcolor{revise_blue}{Random}* \\
            \midrule
            \textcolor{revise_blue}{11}  & \textcolor{revise_blue}{22} & \textcolor{revise_blue}{20, 22, 13, 29} &    \textcolor{revise_blue}{30, 28, 2, 19} \\
            \textcolor{revise_blue}{12} &  \textcolor{revise_blue}{6} &  \textcolor{revise_blue}{5, 13, 27, 11} & \textcolor{revise_blue}{6, 15, 21, 7}\\
            \textcolor{revise_blue}{1}  & \textcolor{revise_blue}{8} & \textcolor{revise_blue}{1, 13, 8, 8} & \textcolor{revise_blue}{18, 15, 14, 13}\\
            \textcolor{revise_blue}{2} &  \textcolor{revise_blue}{5} &  \textcolor{revise_blue}{23, 2, 12, 13} & \textcolor{revise_blue}{24, 23, 13, 15} \\
            \textcolor{revise_blue}{3} &  \textcolor{revise_blue}{18} &  \textcolor{revise_blue}{18, 12, 29, 29} & \textcolor{revise_blue}{28, 7, 26, 4} \\
            \textcolor{revise_blue}{4} &  \textcolor{revise_blue}{2} &  \textcolor{revise_blue}{23, 3, 9, 29} & \textcolor{revise_blue}{11, 24, 11, 15} \\
            \bottomrule
        \end{tabular*}
        \begin{tablenotes}
        \footnotesize
        \item[*] \textcolor{revise_blue}{Four distinct days are chosen for each month, representing electric load, heat load, wind power, and solar power respectively since the methods lack consideration of the interdependent relationship among the four power curves and treat them independently.}
      \end{tablenotes}
    \end{threeparttable}
    \label{tab:scenario}
\end{table}

\textcolor{revise_blue}
{To validate the effectiveness of the proposed time-series scenario generation method, 100 random configuration schemes were selected and evaluated using the model in Section \ref{sect:evaluation model} under four sets of different operational scenarios. These include the typical scenarios generated by the proposed method, the method from \cite{scenario_compare}, randomly selected scenarios, and a full heating season used as a benchmark. The selected typical days are listed in Table \ref{tab:scenario}. It is worth mentioning that the method in \cite{scenario_compare} is selected for comparison due to its consideration of both the variability in renewable energy source power and the heat and electricity load in its operation simulation.}

\textcolor{revise_blue}{
The results, shown in Fig. \ref{fig:scenario_compare}, compare the objective values under these scenarios against the benchmark, with each bar representing the average error across all 100 schemes. As demonstrated, the proposed method achieves lower average error in both objectives, particularly in improving the accuracy of RES power consumption, which is more directly connected to the RES power curves. Lower simulation error allows for the establishment of more complex or larger-scale planning models, which can be solved using the generated typical scenarios. This advantage ensures that the final planning results can perform well in real-world power system planning without costing unaffordable computational resources, thereby maintaining both accuracy and computational feasibility.
}%
\onecolfigure{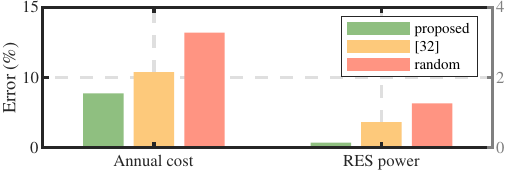}{\textcolor{revise_blue}{The average errors from different scenario selection methods.}}

\begin{figure}[!htbp]
    \centering
    \begin{subfigure}{0.24\textwidth}
        \includegraphics[scale=1]{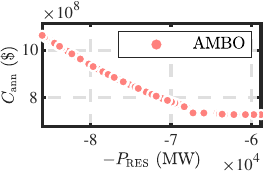}
        \caption{Pareto front of AMBO.}
        \label{fig:PF_noise_AMBO}
    \end{subfigure}
    \begin{subfigure}{0.24\textwidth}
        \includegraphics[scale=1]{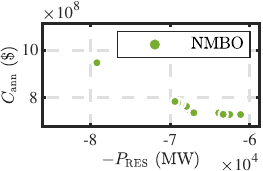}
        \caption{Pareto front of NMBO.}
        \label{fig:PF_noise_NMBO}
    \end{subfigure}
    \begin{subfigure}{0.24\textwidth}
        \includegraphics[scale=1]{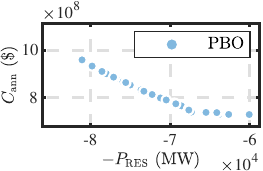}
        \caption{Pareto front of PBO.}
        \label{fig:PF_noise_PBO}
    \end{subfigure}
    \begin{subfigure}{0.24\textwidth}
        \includegraphics[scale=1]{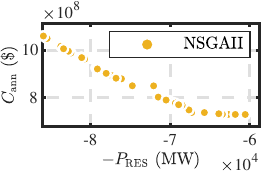}
        \caption{Pareto front of NSGAII.}
        \label{fig:PF_noise_NSGAII}
    \end{subfigure}
    \caption{The Pareto fronts of the five cases under noiseless condition.}
    \label{fig:PF_noiseless}
\end{figure}
\begin{figure}
    \centering
    \includegraphics[scale = 1]{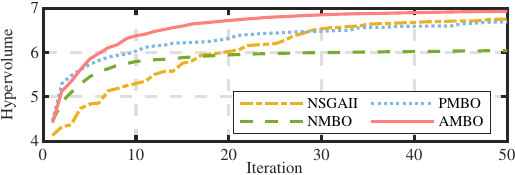}
    \caption{\textcolor{revise_blue}{The hypervolume curves.}}
    \label{fig:HP_typical}
\end{figure}

\textcolor{revise_blue}{As illustrated in Fig. \ref{fig:PF_noiseless}, the Pareto front generated by AMBO is notably more diverse and evenly distributed. This is a direct result of the adaptive determination of the reference point, which allows AMBO to better locate the latent non-dominant points. In contrast, NMBO, which relies on pre-determined parameters, produces a less informative Pareto front. Without the flexibility to adapt to the problem's specific characteristics, demonstrating the limitations of fixed parameters in capturing a comprehensive set of Pareto optimal solutions.} The Pareto fronts of PBO and NSGA-II exhibit clustering, resulting in a suboptimal solution space. From an efficiency standpoint, it is noteworthy that NSGA-II evaluates over four times the number of capacity allocation schemes in each iteration. Combined with the hypervolume curves presented in Fig. \ref{fig:HP_typical}, the Bayesian-based MOO algorithms, particularly AMBO, demonstrate significantly greater sample efficiency.

\onecolfigure{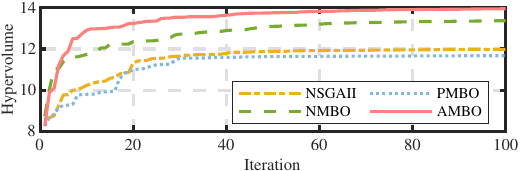}{{\textcolor{revise_blue}{The hypervolume curves calculated by the exam values.}}}
As previously discussed, capacity allocation is planned to address the unpredictable future. However, the outcomes derived from historical scenarios inevitably contain noise. This study employs the sample average approximation (SAA) to establish a benchmark for approximating the actual annual cost and RES power consumption \cite{SAA}. Each capacity allocation scheme undergoes evaluation over a complete heating season, where the corresponding outcomes for each day are summed and averaged to obtain \(C^{\mathrm{SAA}}_{\mathrm{ann}}\) and \(P^{\mathrm{SAA}}_{\mathrm{RES}}\), which are regarded as the actual annual costs of the respective schemes. Detailed formulations are provided below:
\nomenclature[AS]{SAA}{Sample average approximation}
\begin{equation}\label{error_B}
    \left[\begin{aligned}
             & e_\mathrm{B}^\mathrm{ann} \\
             & e_\mathrm{B}^\mathrm{RES}
        \end{aligned}\right]
    = \frac{1}{N_\mathrm{sch}}\sum_{s}^{N_\mathrm{sch}}
    \left[\begin{aligned}
             & \frac{|\mu_{\bm D,1}(\bm x_s) - C^{\mathrm{SAA}}_{\mathrm{ann}}(\bm x_s)|}{C^{\mathrm{SAA}}_{\mathrm{ann}}(\bm x_s)} \\
             & \frac{|\mu_{\bm D,2}(\bm x_s) -P^{\mathrm{SAA}}_\mathrm{RES}(\bm x)|}{P^{\mathrm{SAA}}_\mathrm{RES}(\bm x_s)}
        \end{aligned}\right],
\end{equation}
\begin{equation}\label{error_t}
    \left[\begin{aligned}
             & e_\mathrm{typ}^\mathrm{ann} \\
             & e_\mathrm{typ}^\mathrm{RES}
        \end{aligned}\right]
    = \frac{1}{N_\mathrm{sch}}\sum_{s}^{N_\mathrm{sch}}
    \left[\begin{aligned}
             & \frac{|C^{\mathrm{ob}}_{\mathrm{ann}}(\bm x_s\mid \bm S_\mathrm{typ}) - C^{\mathrm{SAA}}_{\mathrm{ann}}(\bm x_s)|}{C^{\mathrm{SAA}}_{\mathrm{ann}}(\bm x_s)} \\
             & \frac{|P^{\mathrm{ob}}_{\mathrm{RES}}(\bm x_s\mid \bm S_\mathrm{typ}) -P^{\mathrm{SAA}}_\mathrm{RES}(\bm x_s)|}{P^{\mathrm{SAA}}_\mathrm{RES}(\bm x_s)}
        \end{aligned}\right],
\end{equation}
\begin{equation}\label{test_value}
    \left[	\begin{aligned}
             & C_{\mathrm{ann}}(\bm x) \\
             & P_\mathrm{RES}(\bm x)
        \end{aligned}	\right]
    \approx
    \left[ \begin{aligned}
            C^{\mathrm{SAA}}_{\mathrm{ann}}(\bm x) \\
            P_\mathrm{RES}^{\mathrm{SAA}} (\bm x)
        \end{aligned}\right]
    = \frac{1}{N_\mathrm{sce}}\sum_{c=1}^{N_\mathrm{sce}}
    \left[\begin{aligned}
             & C^{\mathrm{ob}}_{\mathrm{ann}}(\bm x\mid \bm P_{c}^\mathrm{sce}) \\
             & P^{\mathrm{ob}}_{\mathrm{RES}}(\bm x\mid \bm P_{c}^\mathrm{sce})
        \end{aligned}\right] ,
\end{equation}
where the error of the Gaussian mean and the original outcome from the evaluation based on the time series scenarios compared with the SAA values of the objectives are denoted by $e_\mathrm{B}^\mathrm{ann/RES}$, $e_\mathrm{typ}^\mathrm{ann/RES}$ respectively.

\onecolfigure{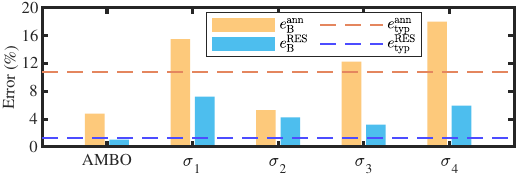}{Average error compared to the SAA outcomes for the two objectives. The bars represent the error of the Gaussian means, while the horizontal lines indicate the error directly from the operational evaluation results.}

With \(\bm \sigma_{n,1}\), \(\bm \sigma_{n,2}\), \(\bm \sigma_{n,3}\), and \(\bm \sigma_{n,4}\) representing four sets of randomly chosen noise standard deviations, the corresponding errors are illustrated in Fig. \ref{fig:error}. Two key conclusions can be drawn from this analysis. First, the proposed time series are effective for evaluating annual costs, as the error from the evaluation outcomes $e_\mathrm{typ}^\mathrm{ann}$ is approximately 1\%, showing no significant discrepancy with $e_\mathrm{B}^\mathrm{ann}$. However, the evaluation of RES consumption proves to be less reliable, with an error exceeding 10\%. This can be attributed to the strong dependence of RES power consumption on the scenarios used for evaluation, making it more challenging to substitute with a smaller time range of time series scenarios.
\textcolor{revise_blue}{The AMBO algorithm demonstrates its capability to effectively handle the noise introduced by simulation deviations, diminishing the error greatly compared with the original outcomes from typical scenarios. This level of accuracy is considered acceptable for practical applications, ensuring the model's reliability without compromising computational efficiency. The adaptive noise modeling in AMBO dynamically adjusts to the simulation deviations, significantly reducing the error without requiring extensive parameter tuning. In contrast, the NMBO algorithm can also achieve a similar error level, but this is contingent on the appropriate tuning of the noise standard deviation \(\bm \sigma_\mathrm{n}\). However, this tuning process is not straightforward and requires substantial computational resources, as there are no established guidelines or systematic methods to determine the optimal values for \(\bm \sigma_\mathrm{n}\). As a result, while NMBO could potentially reach a similar performance to AMBO, it does so at a higher computational cost, highlighting the practical advantage of the adaptive approach in AMBO.}

\subsection{Superiority of the Proposed Approach}
\begin{figure}[!htbp]
    \centering
    \begin{subfigure}{0.49\textwidth}
        \includegraphics[scale=1]{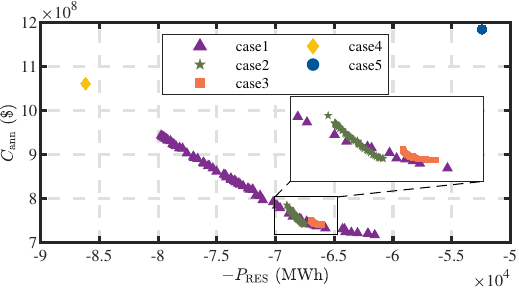}
        \caption{Pareto fronts of all the five cases.}
        \label{fig:PF_mc_full}
    \end{subfigure}
    \begin{subfigure}{0.24\textwidth}
        \includegraphics[scale=1]{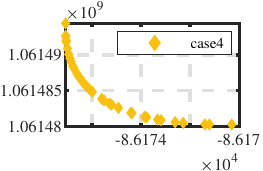}
        \caption{Pareto front of case4.}
        \label{fig:PF_mc_case4}
    \end{subfigure}
    \begin{subfigure}{0.24\textwidth}
        \includegraphics[scale=1]{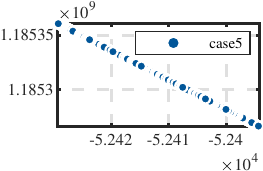}
        \caption{Pareto front of case5.}
        \label{fig:PF_mc_case5}
    \end{subfigure}
    \caption{The Pareto fronts of the five cases.}
    \label{fig:PF_mc}
\end{figure}

In this subsection, five representative cases listed in Table \ref{tab:model_setting} with different planned equipment are analyzed using the AMBO algorithm, and the generated Pareto fronts are illustrated in Fig. \ref{fig:PF_mc}. Notably, Case 1, which is the model selected for this study, offers the most diverse options for planners. In contrast, Cases 2 and 3, which represent the addition of equipment on the generation or consumption side respectively, yield much narrower Pareto fronts. Even more concerning, configurations featuring only an electric boiler or TES result in extremely clustered Pareto fronts, providing planners with limited choices. While planning solely for an electric boiler maximizes RES power consumption, its associated costs may be prohibitively high. Additionally, the performance of TES alone in the proposed model is suboptimal.

\textcolor{revise_blue}{In summary, the results demonstrate that the model performs best when all four kinds of equipment are integrated. This integration is particularly important in cold regions, where effective management of heating and electricity demands is essential. However, the adjustment ability of CHP systems is limited due to the strong coupling between heat and power demands. This highlights the need for a flexible and integrated solution to optimize both heating and electricity production in such regions, ensuring efficient utilization of RES.}
\begin{table}[!htb]
    \footnotesize
    \caption{The Settings of Four Capacity Planning Cases.}
    \centering
    \begin{threeparttable}
        \begin{tabular*}{0.4\textwidth}{@{\extracolsep{\fill}}ccccc}
            \toprule
            Case  & Electric boiler & TES  & Heat pump & CSH \\
            \midrule
            1  & \checkmark & \checkmark & \checkmark & \checkmark \\
            2  & \checkmark & \checkmark & -          & -          \\
            3  & -          & -          & \checkmark & \checkmark \\
            4  & \checkmark & -          & -          & -          \\
            5  & -          & \checkmark & -          & -          \\
            \bottomrule
        \end{tabular*}
    \end{threeparttable}
    \label{tab:model_setting}
\end{table}
\textcolor{revise_blue}{
\subsection{Scalability of the Proposed Approach}
To further assess the scalability of the proposed approach and compare the optimization efficiency of the four MOO algorithms, a simulation was conducted using real-world data from a northern province in China. The data includes information on generator, electric load, heat load, and RES power. The scale of the simulation model was significantly enlarged, now incorporating a total of 53 CHP generators and 32 conventional generators. Additionally, the capacities to be planned for each type of heat source were augmented as well.}
\begin{figure}[!htbp]
    \centering
    \begin{subfigure}{0.24\textwidth}
        \includegraphics[scale=1]{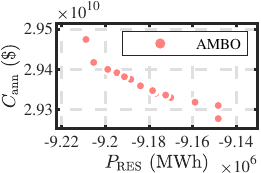}
        \caption{Pareto front of AMBO.}
        \label{fig:larger_PF_AMBO}
    \end{subfigure}
    \begin{subfigure}{0.24\textwidth}
        \includegraphics[scale=1]{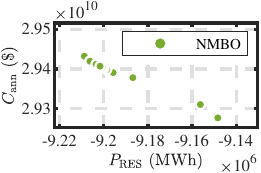}
        \caption{Pareto front of NMBO.}
        \label{fig:larger_PF_NMBO}
    \end{subfigure}
    \begin{subfigure}{0.24\textwidth}
        \includegraphics[scale=1]{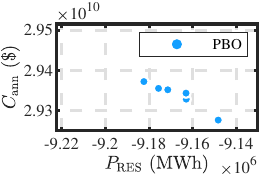}
        \caption{Pareto front of PBO.}
        \label{fig:larger_PF_PBO}
    \end{subfigure}
    \begin{subfigure}{0.24\textwidth}
        \includegraphics[scale=1]{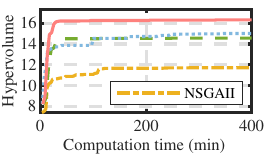}
        \caption{The acquired hypervolume.}
        \label{fig:HP_larger}
    \end{subfigure}
    \caption{\textcolor{revise_blue}{The Pareto fronts and hypervolume comparison with the large-scale model.}}
    \label{fig:PF_larger}
\end{figure}

\textcolor{revise_blue}{
The Pareto fronts of the three Bayesian-based MOO algorithms are presented in Fig. \ref{fig:PF_larger}, along with the hypervolume indicators obtained at different optimization stages. The comparison of the Pareto fronts in Figs. \ref{fig:larger_PF_AMBO}, \ref{fig:larger_PF_NMBO}, and \ref{fig:larger_PF_PBO} clearly shows that the proposed AMBO algorithm generates the most evenly distributed and diverse Pareto front, demonstrating its effectiveness for large-scale models and higher-dimensional optimization problems. Additionally, as illustrated in Fig. \ref{fig:HP_larger}, the AMBO algorithm not only excels in the final hypervolume indicator but also performs better in terms of computational efficiency. In contrast, the NSGA-II algorithm shows lower efficiency due to the significantly greater number of operation simulation runs it requires.
}

\section{Conclusion}
In this study, the proposed AMBO algorithm demonstrates superior performance compared to traditional MOO algorithms by generating a diverse and evenly distributed Pareto front without sacrificing efficiency. These advantages provide planners with a variety of capacity allocation schemes that better accommodate RES power while ensuring reasonable investment choices.
With the aid of the surrogate probability model incorporated in the AMBO, the adverse effects of simulation deviation are effectively mitigated without incurring significant computational costs. Furthermore, the scenarios generated by the proposed time series scenario generation methodology are shown to accurately evaluate a system's annual costs.
The findings also confirm that the collaborative planning of hybrid heat sources from both the generation and consumption sides brings significant benefits to the electricity-heat coupling systems in cold regions.
\bibliography{ref}
\end{document}